\documentclass[conference]{IEEEtran}
\IEEEoverridecommandlockouts
\usepackage{cite}
\usepackage{amsmath,amssymb,amsfonts}
\usepackage{algorithmic}
\usepackage{graphicx}
\usepackage{textcomp}
\usepackage{xcolor}
\usepackage{hyperref}

\def\BibTeX{{\rm B\kern-.05em{\sc i\kern-.025em b}\kern-.08em
    T\kern-.1667em\lower.7ex\hbox{E}\kern-.125emX}}
\begin{document}

\title{Secure synchronization of artificial neural networks used to correct errors in quantum cryptography \break}

\author{
    \IEEEauthorblockN{
    Marcin Niemiec\IEEEauthorrefmark{1},
    Tymoteusz Widlarz\IEEEauthorrefmark{1},
    Miralem Mehic\IEEEauthorrefmark{2}\IEEEauthorrefmark{3}
}

\IEEEauthorblockA{\IEEEauthorrefmark{1}
AGH University of Science and Technology, al. Mickiewicza 30, 30-059 Krakow, Poland
}
\IEEEauthorblockA{\IEEEauthorrefmark{2}
Department of Telecommunications, Faculty of Electrical Engineering, University of Sarajevo, \\
Zmaja od Bosne bb, 71000, Sarajevo, Bosnia and Herzegovina
}
\IEEEauthorblockA{\IEEEauthorrefmark{3}
VSB – Technical University of Ostrava, 17. listopadu 2172/15, 708 00 Ostrava, Czechia
}

\IEEEauthorrefmark{1}niemiec@agh.edu.pl
\IEEEauthorrefmark{2}widlarztymoteusz@gmail.com
\IEEEauthorrefmark{3}miralem.mehic@ieee.org}

\maketitle

\begin{abstract}
Quantum cryptography can provide a very high level of data security. However, a big challenge of this technique is errors in quantum channels. Therefore, error correction methods must be applied in real implementations. An example is error correction based on artificial neural networks. This paper considers the practical aspects of this recently proposed method and analyzes elements which influence security and efficiency. The synchronization process based on mutual learning processes is analyzed in detail. The results allowed us to determine the impact of various parameters. Additionally, the paper describes the recommended number of iterations for different structures of artificial neural networks and various error rates. All this aims to support users in choosing a suitable configuration of neural networks used to correct errors in a secure and efficient way.
\end{abstract}


\section{Introduction}

The emergence and intensive development of the field of quantum computing has put many cryptography algorithms at risk. However, quantum physics also allows to achieve multiple cryptography tasks. One of the most popular is quantum key distribution \cite{ABIDIN2022508}. Unfortunately, quantum communication is not perfect and additional solutions are required to correct any errors after the key distribution in the quantum channel. Artificial neural networks can be utilized to correct these errors \cite{nn_error_correction}. It is a recently proposed solution which provides high level of security and efficiency comparing to other existing error correction methods.

This paper analyzes the impact of different neural networks' parameters on the synchronization process. These parameters influence the number of iterations required as well as the security and efficiency of quantum cryptography. Therefore, it is important to know which neural network scheme should be chosen and which should be avoided. Additionally, the synchronization requires the number of iterations to be specified. Therefore, a recommended number of iterations for a particular multiple neural network's scheme is provided.

The paper is structured as follows. Related work is reviewed in Section 2. Section 3 presents the basics of quantum cryptography, the architecture of the tree parity machine, and error correction using this structure of artificial neural networks. Analysis of synchronization parameters including the recommended number of iterations for typical keys and error rates is described in Section 4. Section 5 concludes the paper.

\section{Related work}

The first quantum key distribution (QKD) protocol, introduced in 1984 by Bennet and Brassard, is BB84 \cite{84_quantum}. This scheme uses the polarization state of a single photon to transmit information. Since then, several other protocols have been presented. One of them is the E91 protocol introduced in 1991 by Ekerd \cite{E91}. It utilizes entangled pairs of photons in the QKD process. However, some errors usually appear during data exchange in the quantum channel. After the initial QKD, there is a specific step: quantum bit error rate (QBER) estimation based on the acquired keys. The QBER value is usually low \cite{QBER_value}. It must to be lower than the chosen threshold used to detect the eavesdropper.

Several methods of correcting error incurred in the quantum key distribution process have been developed. The first described method -- BBBSS -- was proposed in 1992 \cite{BBBSS}. However, the most popular is the Cascade key reconciliation protocol \cite{error_reconcilitation_porownanie}. It is based on multiple random permutations. The Winnow protocol, based on the exchange of parity and Hamming codes, is another method of error correction in the raw key \cite{winnow}. Its main improvement is the reduction of the required communication between both parties. The third most popular error reconciliation scheme is the low density parity check approach. It offers a significant reduction of exchanged information; however, it introduces more computation and memory costs than the Cascade and Winnow protocols \cite{error_reconcilitation_porownanie}.

In 2019, another method of error correction in quantum cryptography was proposed by Niemiec in \cite{nn_error_correction}. The solution uses mutual synchronization of two artificial neural networks (ANN) to correct the errors. The tree parity machine (TPM) is proposed as a neural network used in this approach. It is a well-known structure in cryptography -- the synchronization of two TPMs can be used as a key exchange protocol. TPMs cannot be used as a general method to correct a selected error because it is not possible to predict the final string of bits after the synchronization process. However, it is a desirable feature for shared keys which should be random strings of bits.

\section{Quantum cryptography supported by artificial neural networks}

Symmetric cryptography uses a single key to encrypt and decrypt secret messages. Let's assume that Alice and Bob, the two characters used in describing cryptography protocols, are using symmetric encryption. The goal is to send information from Alice to Bob in a way that provides confidentiality. To achieve this, Alice and Bob need to agree on a shared secret key. Alice encrypts confidential data using the previously chosen key and Bob decrypts it using the same key. The same key is applied to encrypt and decrypt the information, hence the name: symmetric-key encryption. It is worth mentioning only the one-time-pad symmetric scheme has been proven secure but it requires a key not smaller than the message being sent.

In general, symmetric-key encryption algorithms -- for example the Advanced Encryption Standard (AES) \cite{symetric_crypto_book} -- perform better than asymmetric-key algorithms \cite{performance}. However, symmetric-key algorithms have an important disadvantage compared to asymmetric-key schemes. In the symmetric key encryption scheme, the key needs to be safely distributed or established between Alice and Bob \cite{key}.
The symmetric key can be exchanged in a number of ways, including via a trusted third party or by direct exchange between involved parties. However, both methods introduce some vulnerabilities, including passive scanning of network traffic. A method where the eavesdropper can be easily detected uses quantum mechanics to establish keys between Alice and Bob. It is called the quantum key distribution protocol.

\subsection{Quantum key distribution}

Quantum mechanics allows for secure key distribution\footnote{In fact, a key is not distributed but negotiated. However, the term 'distribution' is consistently used in this paper to be consistent with the commonly accepted name of the technique.} among network users. Two main principles are the core of the security of QKD: an unknown quantum state cannot be copied \cite{no_cloning}, and the quantum state cannot be estimated without disturbing it. One of the most popular QKD protocols which uses those principles is the BB84 scheme \cite{84_quantum}.

The BB84 protocol uses photons with two polarization bases: rectilinear or diagonal. Alice encodes a string of bits using photons on a randomly chosen basis. After that, all the photons are sent through a quantum channel. Bob randomly chooses a basis for each photon to decode the binary $0$ or $1$. Alice and Bob's bases are compared through a public communication channel. Each bit where both parties chose the same basis should be the same. However, when Bob measures the photon in a different basis than Alice, this bit is rejected. The remaining bits are the same for both parties and can be considered as a symmetric key. Next, the error estimation is performed. Randomly chosen parts of the keys between Alice and Bob are compared to compute  the QBER value. If the comparison results in a high error rate, it means that the eavesdropper (Eve) is trying to gain information about the exchanged photons. However, the quantum channel is not perfect, and errors are usually detected due to disturbance, noise in the detectors or other elements. The number of errors introduced by the quantum channel's imperfections must be considered while deciding the maximum acceptable error rate.

The differences between Alice and Bob's keys need to be corrected. Several error correction methods are known. BBBSS is the earliest scheme proposed in \cite{BBBSS}. It is mainly based on parity checks. The most popular method is the Cascade protocol \cite{cascade}. It is an improved version of BBBSS and requires less information to be sent between Alice and Bob through the public channel.
The Cascade protocol and its predecessor are based on multiple parity checks. The basic idea is that the keys are divided into blocks of a fixed size. The number of bits in each block depends on the previously calculated QBER value. Alice and Bob compare the parities of each block to allow them to find an odd number of errors. If errors are detected in a given block, it is split into two. The process is repeated recursively for each block until all errors are corrected. It concludes a single iteration after which Alice and Bob have keys with an even number of errors or without any errors. Before performing the following iterations, the keys are scrambled, and the size of the block is increased. The number of iterations is predetermined. As a result of this process, Alice and Bob should have the same keys. However, it is not always the case. A number of iterations or block sizes can be chosen incorrectly and cause failure in error correction. Additionally, the algorithm performs multiple parity checks over the public channel, which can be intercepted by an eavesdropper (Eve). As a result, Eve can construct a partial key. Alice and Bob should discard parts of their keys to increase the lost security. This reduces the performance of this method since the confidential keys must be shortened in the process. Another error reconciliation method is based on mutual synchronization of artificial neural networks.

\subsection{Tree parity machine}

An artificial neural network (ANN) is a computing system inspired by biological neural networks \cite{ann}. ANNs are used to recognize patterns and in many other solutions in the fields of machine learning. ANNs consist of multiple connected nodes (artificial neurons), with each neuron representing a mathematical function \cite{neuro}. These nodes are divided into three types of layers: the first (input) layer, at least one hidden layer, and the output layer. The connections between neurons in each layer can be characterized by weights.

\begin{figure*}[htbp]
    \centering
    \includegraphics[width=10cm, keepaspectratio]{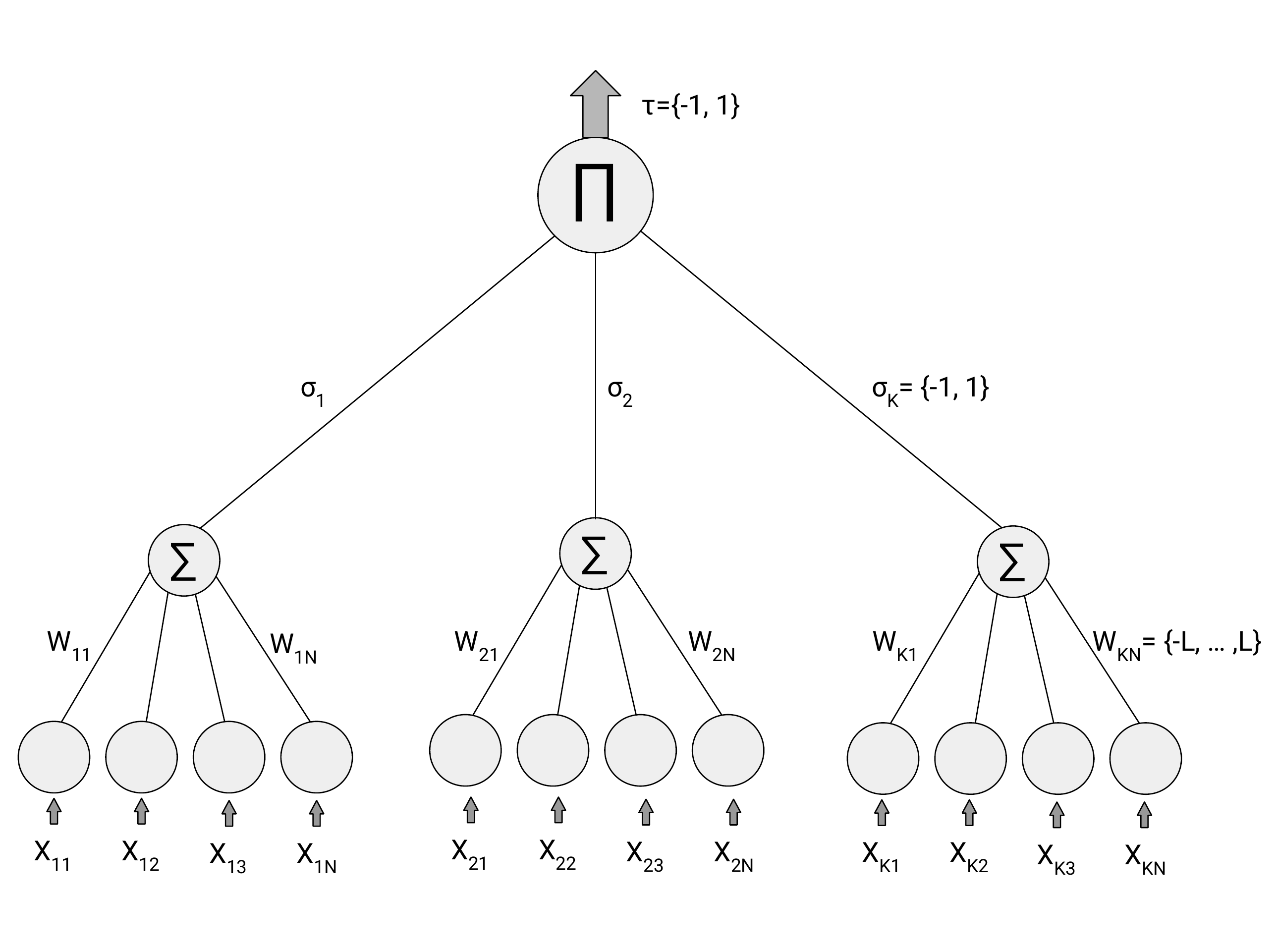}
    \caption{Model of tree parity machine.} 
    \label{fig:TPM}
\end{figure*}

In cryptography, the most commonly used neural network is the tree parity machine (TPM) \cite{neuron3}. A scheme of this model is presented in Fig. \ref{fig:TPM}.
There are $K\times N$ input neurons, divided into $K$ groups. There is a single hidden layer with $K$ nodes. Each of these nodes has $N$ inputs. The TPM has a single output neuron. The connections between input neurons and hidden layer neurons are described by weights $W$ -- integers in the range [$-L$, $L$], thus $L$ is the maximum and $-L$ is the minimum weight value.
The values of $\sigma$ characterize the connections between the hidden layer neurons and an output neuron. The output value of the TPM is described by $\tau$.

The value of $\sigma$ is calculated using the following formulas:
\begin{equation}
    \sigma_k = sgn(\sum_{n=1}^{N} x_{kn} * w_{kn})
\end{equation}
\begin{equation}
    sgn(z) = \begin{cases}
      -1 & z\leq0\\
      1 & z > 0\\
      \end{cases}
\end{equation}
Due to the usage of the presented signum function, $\sigma$ can take two values: $1$ or $-1$. The output value of TPM is calculated as:
\begin{equation}
    \tau = \prod_{k=1}^K \sigma_k
\end{equation}
This neural network has two possible outcomes: $1$ or $-1$.

For the TPM structure, multiple learning algorithms are proposed. Most popular are Hebbian, anti-Hebbian, and random walk. The leading is the Hebbian rule \cite{hebbian}. The Hebbian algorithm updates ANN weights in the following manner:

\begin{equation}
    w^*_{kn} = v_L(w_{kn} + x_{kn} * \sigma_k * \theta(\sigma_k, \tau))
\end{equation}
where $\theta$ limits the impact of hidden layer neurons whose value was different than $\tau$:

\begin{equation}
    \theta(\sigma_k, \tau) = \begin{cases}
                                0 & \text{if } \sigma_k \neq \tau\\
                                1 & \text{if } \sigma_k = \tau
                            \end{cases}
\end{equation}
The $v_L$ function makes sure that the new weights are kept within the [$-L$, $L$] range:

\begin{equation}
    v_L(z) = \begin{cases}
                -L & \text{if } z \leq -L\\
                z & \text{if } -L < z < L\\
                L & \text{if } z \geq L
            \end{cases}
\end{equation}

The TPM structure allows for mutual learning of the two neural networks \cite{mutual_learning}, primarily based on updating weights only when the outputs from both neural networks are the same. The input values are random and the same for both Alice and Bob's TPMs. Inputs are updated in each iteration. The security of this process relies on the fact that cooperating TPMs can achieve convergence significantly faster than Eve's machine, which can update weights less frequently.
The TPM is most commonly used in cryptography to exchange a secret key. This usage is defined as neural cryptography \cite{neural_cryptography}. Alice and Bob mutually synchronize their TPMs to achieve the same weights. After the synchronization process, these weights provide a secure symmetric key.

\subsection{Error correction based on TPMs}

TPMs can be utilized during the error correction process in quantum cryptography \cite{nn_error_correction}. The neural network's task is to correct all errors to achieve the same string of confidential bits at both endpoints. 
Firstly, Alice and Bob prepare their TPMs. The number of neurons in the hidden layer ($K$) and the number of input neurons ($N$) is determined by Alice and passed on to Bob. The value $L$ must also be agreed between the users. The keys achieved using the QKD protocol are changed into integer values in the range [$-L$, $L$]. These values are used in the appropriate TPMs as weights between neurons in the input layer and the hidden layer. Since Alice's string of bits is similar to Bob's (QBER is usually not high), the weights in the created TPMs are almost synchronized. At this point, Alice and Bob have constructed TPMs with the same structure but with a few differences in the weight values.

After establishing the TPM structure and changing bits to weights, the synchronization process starts. It consists of multiple iterations, repeated until common weights are achieved between Alice and Bob. A single iteration starts from Alice choosing the input string and computing the result using the TPM. After that, the generated input string is passed on to Bob, who computes the output of his TPM using the received input. Then, the results are compared. If the outputs of both TPMs match, the weights can be updated. Otherwise, the process is repeated with a different input string.

After an appropriate number of iterations, the TPMs are synchronized and Alice and Bob can change the weights back into a string of bits. The resulting bits are the same. However, the privacy amplification process after error correction is still recommended \cite{amplification}.
The reduction of the key protecting Alice and Bob from information leakage is defined as \cite{nn_error_correction}:
\begin{equation}
    Z = log_{2L+1}2^i
\end{equation}
where $i$ is the number of TPM iterations.

This usage of TPM is safer than the neural cryptography solution, because weights are similar before the synchronization. Therefore, significantly fewer iterations are required to achieve convergence than the randomly initialized weights in key establishing algorithms.
It is worth mentioning this method of error correction is characterized by high efficiency, e.g. requires approximately 30\% less iterations than Cascade algorithm \cite{nn_error_correction}.

\section{Analysis of the synchronization process}

The crucial decision regarding the error detection approach based on TPMs is the number of iterations during the synchronization process. This value should be as low as possible for security reasons. However, it cannot be too low, since neural networks will not be able to correct all errors in the key otherwise. It is the user's responsibility to select the appropriate value for the error correction. The main objective of the analysis is to determine the impact of various neural network parameters on the synchronization process.
Another goal is to provide a recommended number of iterations for users.

\subsection{Testbed}

The experiments require an application to simulate the error correction process based on artificial neural networks. The application for correcting errors arising in quantum key distribution was written in Python and uses the NumPy package -- a library for scientific computing which provides fast operations on arrays required by the TPM. The functions provided by NumPy satisfy all necessary calculations to achieve neural network convergence. Synchronization of TPMs is performed over sockets to allow real-world usage of this tool. The Hebbian learning algorithm for updating weights is used.

The developed application makes it possible to correct errors in the keys using quantum key distribution protocols. The users are also able to correct simulated keys with the chosen error rate. It helps if users do not have strings of bits created by a real QKD system. An important feature of the tool is its ability to select neural network parameters. The user can personalize the synchronization process, starting from the key length and error rate. The least sufficient number of bits was used for translation into a single integer (values of the weights must be in the range [$-L$, $L$]). It was demonstrated that the number of hidden neurons and the number of inputs depend on the chosen key length and $L$ value. Therefore, users need to select these parameters taking into account the requirements and needs.

During the experiments the minimum number of returned required iterations for a single TPM configuration was set to $200$. The maximum number of iterations was limited to $1000$. Additionally, the maximum number of retries in a single iteration was limited to $10$ to speed up the simulation process.
Finally, $1880$ different scenarios were analyzed. All possible TPM configurations for key lengths varying between $100$ and $700$ with a $100$ bit step are available. Moreover, the data is available for other keys with lengths varying between $128$ and $352$ with an $8$ bit step.
Between $350$ and $500$ synchronizations were performed for each TPM. It was assumed that this number of iterations is sufficient to achieve convergence.

\subsection{Recommended number of iterations}

To obtain the recommended number of iterations of TPMs for successful error correction, the sum of means and standard deviations of the results was calculated. The median and variance values were calculated as well for comparison. The full results are available online\footnote{Recommended numbers of iterations for $1880$ different scenarios -- TPM structures and QBER values -- are available from: \url{http://kt.agh.edu.pl/~niemiec/ICC-2023}  This is mainly based on possible key lengths which vary between $128$ and $500$ bits with $4$ bit steps. Additionally, keys with lengths between $500$ and $700$ with $100$ bit steps are included.}.
The selected part -- the neural network configurations where the key length equals $256$ bits with the recommended number of iterations -- is presented in Tab. \ref{tab:iterations}.

\begin{table}[htbp]
    \caption{Recommended number of iterations for TPMs generated for 256 bit keys}
    \centering
    \begin{tabular}{|p{1.1cm}|p{0.8cm}|p{1.3cm}|p{1.1cm}|p{1.7cm}|}
    \hline
        \textbf{Weights range} \newline \{$-L$, $L$\} & 
        \textbf{QBER} [$\%$] & 
        \textbf{Number of inputs to a single hidden neuron} [\emph{N}] & 
        \textbf{Number of hidden neurons} [\emph{K}] & 
        \textbf{Recommended number of iterations}  \\ \hline
        2	& 1	& 2	& 43  & 154  \\ \hline
        2 & 1 & 43 & 2  & 51   \\ \hline
        2 & 2 & 2  & 43 & 179  \\ \hline
        2 & 2 & 43 & 2  & 59   \\ \hline
        2 & 2 & 86 & 1  & 24   \\ \hline
        2 & 3 & 2  & 43 & 188  \\ \hline
        2 & 3 & 43 & 2  & 64   \\ \hline
        2 & 3 & 86 & 1  & 25   \\ \hline
        3 & 1 & 2  & 43 & 218  \\ \hline
        3 & 1 & 43 & 2  & 71   \\ \hline
        3	& 1	& 86 & 1  &	33   \\ \hline
        3 & 2 & 2  & 43 & 309  \\ \hline
        3 & 2 & 43 & 2  & 94   \\ \hline
        3	& 2	& 86 & 1  &	39   \\ \hline
        3 & 3 & 2  & 43 & 325  \\ \hline
        3 & 3 & 43 & 2  & 97   \\ \hline
        3 & 3 & 86 & 1  & 40   \\ \hline
        4 & 1 & 2  & 32 & 450  \\ \hline
        4 & 1 & 4  & 16 & 496  \\ \hline
        4 & 1 & 8  & 8  & 301  \\ \hline
        4 & 1 & 16 & 4  & 176  \\ \hline
        4 & 1 & 32 & 2  & 125  \\ \hline
        4 & 2 & 2  & 32 & 554  \\ \hline
        4 & 2 & 4  & 16 & 701  \\ \hline
        4 & 2 & 8  & 8  & 483  \\ \hline
        4 & 2 & 16 & 4  & 264  \\ \hline
        4 & 2 & 32 & 2  & 152  \\ \hline
        4 & 3 & 2  & 32 & 609  \\ \hline
        4 & 3 & 4  & 16 & 772  \\ \hline
        4 & 3 & 8  & 8  & 542  \\ \hline
        4 & 3 & 16 & 4  & 302  \\ \hline
        4 & 3 & 32 & 2  & 164  \\ \hline
    \end{tabular}
    \label{tab:iterations}
\end{table}

\begin{figure}[ht]
    \centering
    \includegraphics[width=9cm, keepaspectratio]{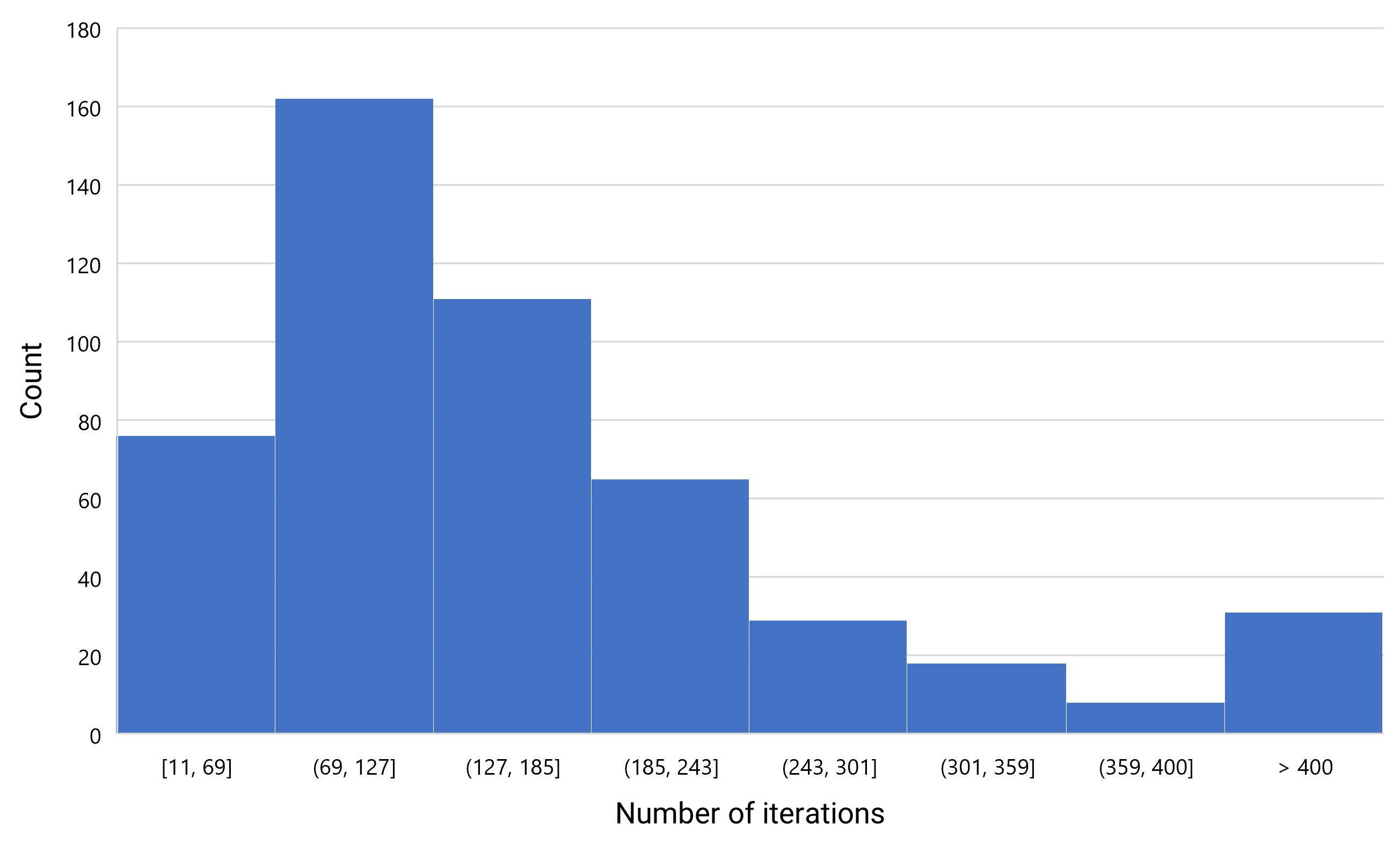}
    \caption{Histogram for number of iterations (TPM with a $256$ bit key, $N=16$, $K=4$, $L=4$, $QBER=3\%$).}
\label{fig:256_histogram}
\end{figure} 

Fig. \ref{fig:256_histogram} shows the histogram of data gathered for a single neural network configuration. The distribution is right-skewed. The mean value is greater than the median. It is a common characteristic for other tested TPM configurations. If the distribution is not positively skewed, it is symmetrical. The recommended number of iterations for the presented configuration, according to Tab. \ref{tab:iterations}, equals $302$. It is based on the sum of the mean and standard deviation values. For all presented TPM configurations, this sum gives an $84\%$ chance of successful synchronization, assuming a normal distribution of results. For the right-skewed distribution, similar to the one presented in Fig. \ref{fig:256_histogram}, the probability of success is higher. The $85$-th percentile for the given set is equal to $276$ -- less than the proposed value. In this case, after choosing the suggested number of iterations the user has more than an $88\%$ chance of success.

Knowing the lowest required number of iterations is important because it reduces the risk of a successful attack by Eve. The attacker could create independent TPMs and try to synchronize one of them with Alice or Bob's machine. The recommended number of iterations increases the security of this solution because Alice and Bob require far fewer iterations to synchronize, compared to Alice (or Bob) and Eve synchronizing using random weights.

\subsection{Impact of TPM structures}

\begin{figure}[ht]
    \centering
    \includegraphics[width=9cm, keepaspectratio]{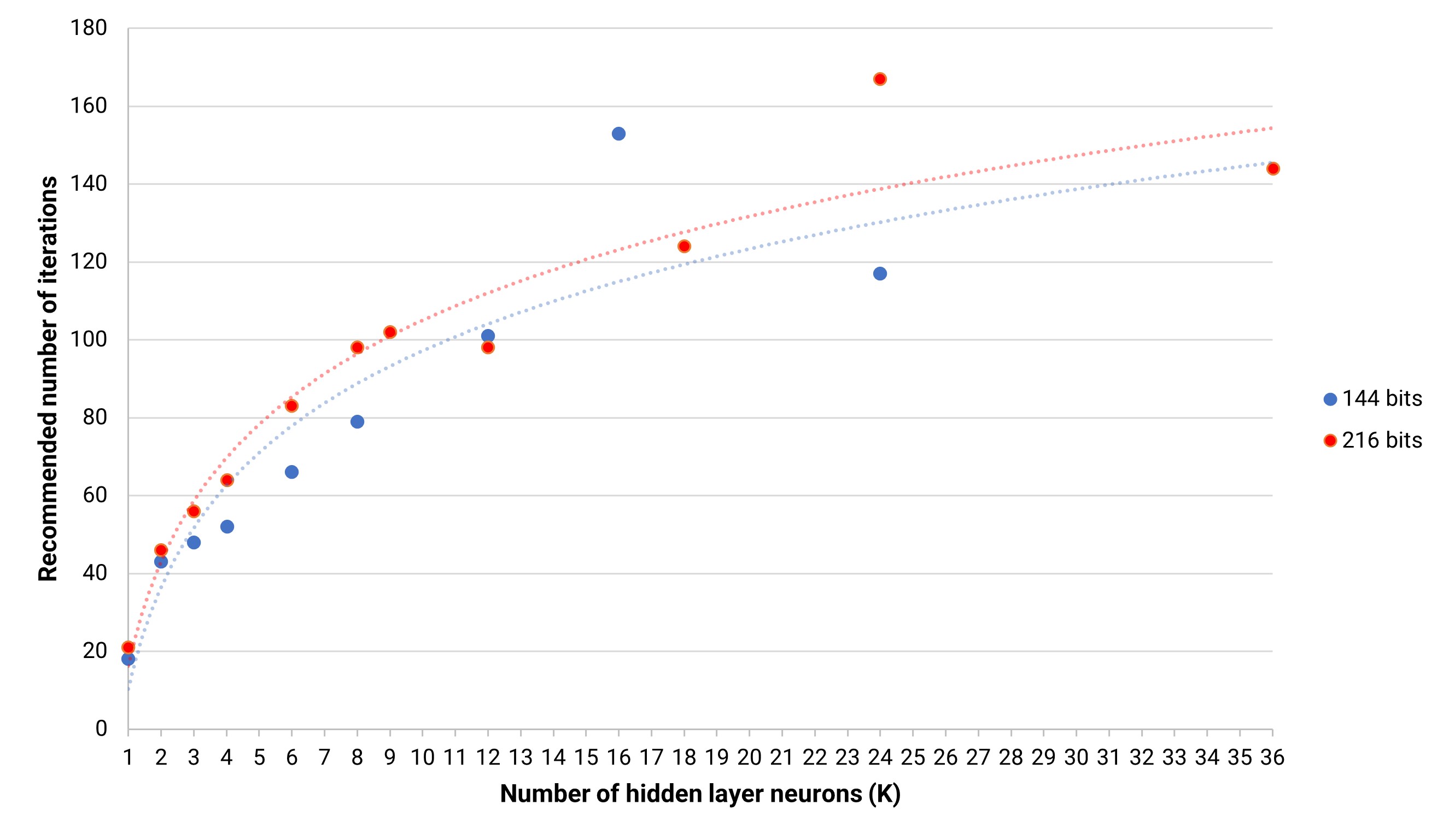}
    \caption{\label{fig:K_impact} Number of iterations for TPMs with $144$ and $216$ bit keys for different $K$ value.}
\end{figure}

The results of simulations allow us to analyze how TPM structures affect the number of required iterations during the synchronization process. 
Fig. \ref{fig:K_impact} shows the number of required iterations depending on the $K$ and $N$ parameters. It shows two different TPM configurations: one with a $144$ bit key and another with a $216$ bit key. These configurations were chosen due to having a similar number of possible $K$ and $N$ pairs. For a given key length, $L$ value and error rate there is a limited number of possible $N$ and $K$ values. The $K$ value changes in inverse proportion to the $N$ value. As presented in Fig. \ref{fig:K_impact} the speed of the TPM synchronization process depends on the neural network structure ($N$ and $K$ values). The number of required iterations increases alongside the higher number of neurons in the hidden layer ($K$). The trend is similar for both presented TPMs. After achieving a certain threshold, the number of recommended iterations increases slowly. The results fit the logarithmic trend line. It means that after a certain $K$ value, increasing this parameter further does not affect the synchronization speed as much as under a certain threshold.

\begin{figure}[ht]
    \centering
    \includegraphics[width=9cm, keepaspectratio]{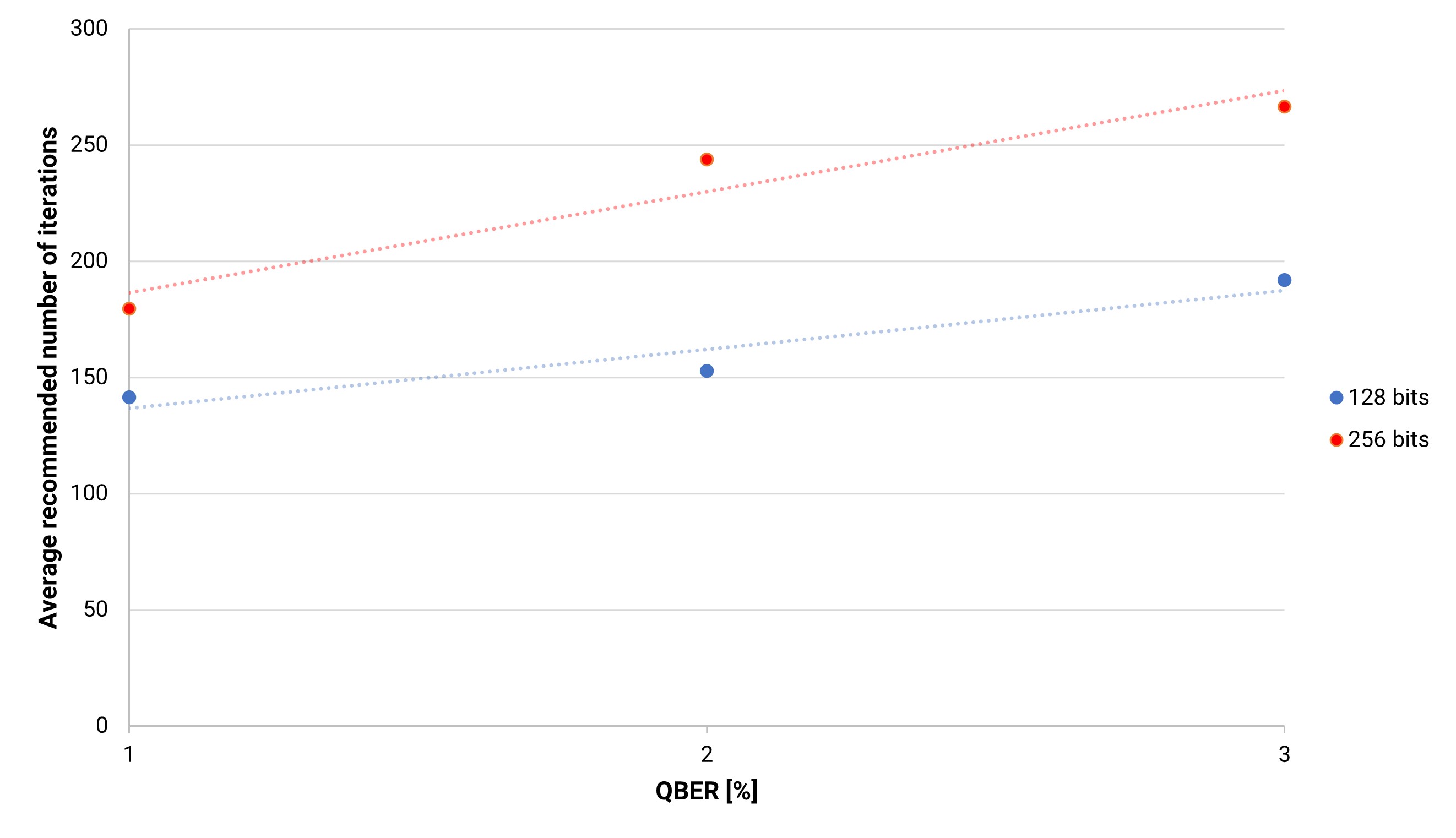}
    \caption{\label{fig:QBER_impact} Number of iterations for TPMs with $128$ and $256$ bit keys depended on the QBER.}
\end{figure}

Other configurations of the selected TPMs were studied based on the increasing error rate of the keys. Two configurations with $128$ and $256$ bit keys were tested. The average of every possible configuration of the recommended number of iterations was calculated for different QBER values. The results are presented in Fig. \ref{fig:QBER_impact}. This confirms that a greater number of errors results in a higher average number of recommended iterations. It confirms the applicability of TPMs to correct errors emerging in quantum key distribution, where the error rate should not be higher than a few percent. Therefore, the eavesdropper needs more iterations to synchronize its TPM.

\begin{figure}[ht]
    \centering
    \includegraphics[width=9cm, keepaspectratio]{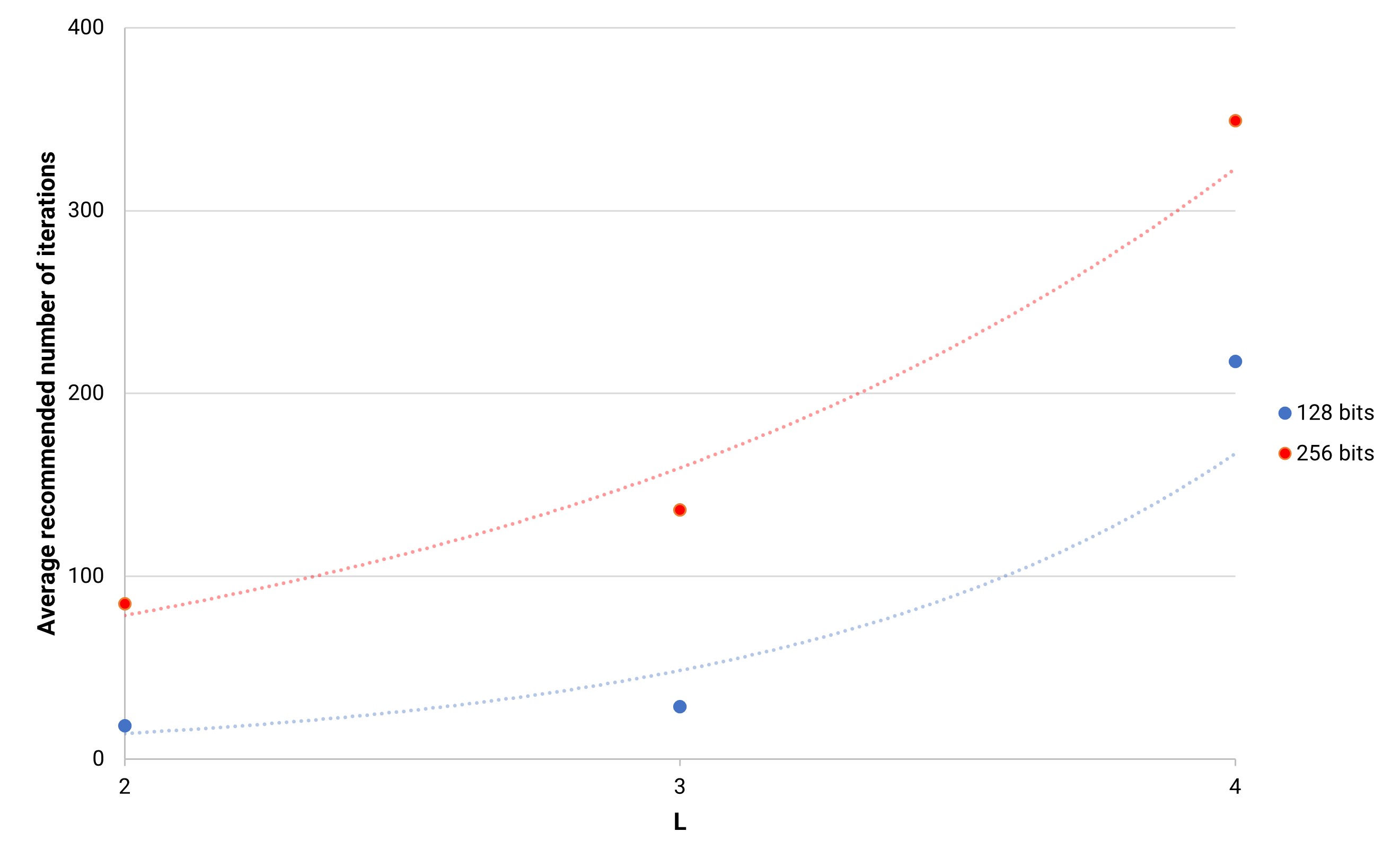}
    \caption{\label{fig:L_impact} Number of iterations for TPMs with $128$ and $256$ bit keys dependent on the $L$ value.}
\end{figure}

Additionally, it was verified that value $L$ has an exponential impact on the average recommended number of iterations. The data was gathered using a similar approach to the study with the impact of QBER. The average recommended number of iterations of each configuration for a given $L$ was calculated. Fig. \ref{fig:L_impact} shows the exponential trend line. It is worth mentioning that the impact of $L$ value on the synchronization time is significant.

It is the user's responsibility to choose the best possible configuration for a given key length and QBER value. The analysis shows that the $L$ value should be chosen carefully since it exponentially affects the required number of iterations. Additionally, the choice of the $K$ value should be made with caution due to its logarithmic impact on the number of iterations.

\section{Summary}

The analysis of the TPM synchronization process used for error correction purposes was presented in this paper. It shows that the parameters of the TPM structure have an impact on the synchronization time and security of this error correction method. However, different parameters of artificial neural networks have different effects. Therefore, users should be aware of how to choose the configuration of neural networks used to correct errors in a secure and efficient way. One of the deciding factors which need to be selected is the number of iterations. The paper describes the recommended number of iterations for different TPM structures and QBER values to assist users in this step. The numbers recommended by the authors are as low as possible but with a high probability of successful synchronization to ensure secure and efficient error correction based on artificial neural networks.

\section*{Acknowledgment}
This work was supported by the ECHO project which has received funding from the European Union’s Horizon 2020 research and innovation programme under the grant agreement no. 830943.

\bibliographystyle{IEEEtran}
\bibliography{IEEEabrv,ref}

\begin{thebibliography}{10}
\providecommand{\url}[1]{#1}
\csname url@samestyle\endcsname
\providecommand{\newblock}{\relax}
\providecommand{\bibinfo}[2]{#2}
\providecommand{\BIBentrySTDinterwordspacing}{\spaceskip=0pt\relax}
\providecommand{\BIBentryALTinterwordstretchfactor}{4}
\providecommand{\BIBentryALTinterwordspacing}{\spaceskip=\fontdimen2\font plus
\BIBentryALTinterwordstretchfactor\fontdimen3\font minus
  \fontdimen4\font\relax}
\providecommand{\BIBforeignlanguage}[2]{{%
\expandafter\ifx\csname l@#1\endcsname\relax
\typeout{** WARNING: IEEEtran.bst: No hyphenation pattern has been}%
\typeout{** loaded for the language `#1'. Using the pattern for}%
\typeout{** the default language instead.}%
\else
\language=\csname l@#1\endcsname
\fi
#2}}
\providecommand{\BIBdecl}{\relax}
\BIBdecl

\bibitem{ABIDIN2022508}
S.~Abidin, A.~Swami, E.~Ramirez-Asís, J.~Alvarado-Tolentino, R.~K. Maurya, and
  N.~Hussain, ``Quantum cryptography technique: A way to improve security
  challenges in mobile cloud computing (mcc),'' \emph{Materials Today:
  Proceedings}, vol.~51, pp. 508--514, 2022.

\bibitem{nn_error_correction}
M.~Niemiec, ``Error correction in quantum cryptography based on artificial
  neural networks,'' \emph{Quantum Information Processing}, 2019.

\bibitem{84_quantum}
C.~Bennett and G.~Brassard, ``Quantum cryptography: Public key distribution and
  coin tossing,'' \emph{Theoretical Computer Science - TCS}, pp. 175--179,
  1984.

\bibitem{E91}
A.~Ekert, ``{Quantum cryptography based on Bell's theorem},'' \emph{Phys. Rev.
  Lett.}, pp. 661--663, 1991.

\bibitem{QBER_value}
M.~Khodr, ``Evaluations of quantum bit error rate using the three stage
  multiphoton protocol,'' \emph{2017 International Conference on Electrical and
  Computing Technologies and Applications (ICECTA)}, pp. 1--4, 2017.

\bibitem{BBBSS}
C.~Bennett, F.~Bessette, G.~Brassard, L.~Salvail, and J.~Smolin, ``Experimental
  quantum cryptography,'' \emph{Journal of Cryptology}, pp. 3--28, 1992.

\bibitem{error_reconcilitation_porownanie}
M.~Mehic, M.~Niemiec, H.~Siljak, and M.~Voznak, ``Error reconciliation in
  quantum key distribution protocols,'' \emph{Reversible Computation: Extending
  Horizons of Computing: Selected Results of the COST Action IC1405}, pp.
  222--236, 2020.

\bibitem{winnow}
W.~T. Buttler, S.~K. Lamoreaux, J.~R. Torgerson, G.~H. Nickel, C.~H. Donahue,
  and C.~G. Peterson, ``Fast, efficient error reconciliation for quantum
  cryptography,'' \emph{Phys. Rev. A}, 2003.

\bibitem{symetric_crypto_book}
H.~Delfs and H.~Knebl, ``Symmetric-key encryption,'' \emph{Introduction to
  Cryptography: Principles and Applications}, pp. 11--31, 2007.

\bibitem{performance}
M.~Panda, ``Performance analysis of encryption algorithms for security,''
  \emph{2016 International Conference on Signal Processing, Communication,
  Power and Embedded System (SCOPES)}, pp. 278--284, 2016.

\bibitem{key}
M.~Umaparvathi and D.~K. Varughese, ``Evaluation of symmetric encryption
  algorithms for manets,'' \emph{2010 IEEE International Conference on
  Computational Intelligence and Computing Research}, pp. 1--3, 2010.

\bibitem{no_cloning}
W.~K. Wootters and W.~H. Zurek, ``A single quantum cannot be cloned,''
  \emph{Nature}, pp. 802--803, 1982.

\bibitem{cascade}
G.~Brassard and L.~Salvail, ``Secret-key reconciliation by public discussion,''
  \emph{Advances in Cryptology}, pp. 410--423, 1994.

\bibitem{ann}
J.~Hopfield, ``Artificial neural networks,'' \emph{IEEE Circuits and Devices
  Magazine}, pp. 3--10, 1988.

\bibitem{neuro}
P.~P. Hadke and S.~G. Kale, ``Use of neural networks in cryptography: A
  review,'' in \emph{2016 World Conference on Futuristic Trends in Research and
  Innovation for Social Welfare (Startup Conclave)}, 2016, pp. 1--4.

\bibitem{neuron3}
A.~Sarkar, ``Secure exchange of information using artificial intelligence and
  chaotic system guided neural synchronization,'' \emph{Multimedia Tools and
  Applications}, vol.~80, pp. 1--31, 05 2021.

\bibitem{hebbian}
M.~Aleksandrov and Y.~Bashkov, ``Factors affecting synchronization time of tree
  parity machines in cryptography,'' \emph{2020 IEEE 2nd International
  Conference on Advanced Trends in Information Theory (ATIT)}, pp. 108--112,
  2020.

\bibitem{mutual_learning}
R.~Metzler, W.~Kinzel, and I.~Kanter, ``Interacting neural networks,''
  \emph{Phys. Rev. E}, pp. 2555--2565, 2000.

\bibitem{neural_cryptography}
W.~Kinzel and I.~Kanter, ``Neural cryptography,'' \emph{Proceedings of the 9th
  International Conference on Neural Information Processing}, pp. 1351--1354,
  2002.

\bibitem{amplification}
C.~Bennett, G.~Brassard, and J.~Robert, ``Privacy amplification by public
  discussion,'' \emph{SIAM J. Comput.}, p. 210–229, 1988.

\end{thebibliography}

\end{document}